# Mechanical force involved multiple fields switching of both local ferroelectric and magnetic domain in a $Bi_5Ti_3FeO_{15}$ thin film

Tingting Jia[1], Hideo Kimura[1], Zhenxiang Cheng[2], Hongyang Zhao[3], Yoon-Hyun Kim[1], Minoru Osada[1], Takao Matsumoto[4], Naoya Shibata[4], Yuichi Ikuhara[4]

[1] National Institute for Materials Science, 1-2-1 Sengen, Tsukuba, Ibaraki 305-0047, Japan
[2] Institute for Superconducting & Electronic Materials, University of Wollongong, Innovation Campus, North Wollongong, NSW 2500, Australia
[3] Department of Materials Science and Engineering, Wuhan Institute of Technology, Wuhan 430073, China
[4] Institute of Engineering Innovation, School of Engineering, The University of Tokyo, 2-11-16 Yayoi, Bunkyo-ku, Tokyo 113-8656, Japan

## Abstract

Multiferroics have received intense attention due to their great application potential in multi-state information storage devices and new types of sensors. Coupling among ferroic orders such as ferroelectricity, (anti-)ferromagnetism, ferroelasticity, etc. will enable dynamic interaction between these ordering parameters. Direct visualization of such coupling behaviour in single phase multiferroic materials is highly desirable for both applications and fundamental study. Manipulation of both ferroelectric and magnetic domains of $Bi_5Ti_3FeO_{15}$ thin film using electric field and external mechanical force is reported, which confirms the magnetoelectric coupling in $Bi_5Ti_3FeO_{15}$, indicates the electric and magnetic orders are coupled through ferroelasticity. Due to the anisotropic relaxation of ferroelastic strain, the back-switching of out-of-plane electric domains is not as obvious as in-plane. An inevitable destabilization of the coupling between elastic and magnetic ordering happens because of the elastic strain relaxation, which result in a subsequent decay of magnetic domain switching. Mechanical force applied on the surface of $Bi_5Ti_3FeO_{15}$ film generates by an AFM tip will effectively drive a transition of the local ferroelastic strain state, reverse both the polarization and magnetization in a way similar to an electric field. Current work provides a framework for exploring cross-coupling among multiple orders and potential for developing novel nanoscale functional devices.

Keywords: multiferroic materials, domain, scanning probe microscopy, magnetoelectric coupling and thin film.

**INTRODUCTION**

Multiferroic materials have been drawing extensive attention worldwide because they simultaneously possess ferroelectric, magnetic, and/or ferroelastic properties. The coupling between these properties enables the dynamic manipulation of one ordering parameter by another, which is promising for broad applications in sensing, actuation, memory, etc.[1, 2, 3] Numerous efforts have been made to investigate the coupling phenomena, especially the coupling between electric and magnetic orderings that would account for the magnetoelectric (ME) effect in multiferroic materials, due to the rich fundamental physics and exciting application potential in multiply controlled devices. [3,4] Recently, the development of scanning probe microscopy (SPM) techniques has enabled us to not only directly visualize, but also switch both ferroelectric domains and magnetic domains. [5, 6] Therefore, the direct visualization of the coupling between magnetic domains and ferroelectric domains has been made possible and recently has become a powerful tool for ME studies. Zhao et al. observed antiferromagnetic domain switching induced by ferroelectric polarization switching in $BiFeO_3$ thin films,[7] while Keeney et al. demonstrated that ferroelectric domain polarization switching could be induced by an applied magnetic field in Aurivillius-phase multiferroic thin films.[8] These works have demonstrated the powerful capability of SPM technology in the study of ME coupling. Furthermore, investigations on the nanoscale ferroelectric domain switching under mechanical force loaded by a SPM tip is drawing increasing attention.[9, 10] Recently, mechanical-force-induced ferroelectric domain switching in multiferroics has been reported by several groups.[11, 12, 13] It is natural to ask whether such an external mechanical force could switch two or all three types of ferroic domains (ferroelectric, ferroelastic and magnetic domains) in multiferroic materials. Finding the answer in multiferroics is of fundamental and practical importance for understanding their quasistatic and dynamic behavior in multiply controlled devices applications.

Zhao et al. found that strong magnetoelectric coupling in pure $Bi_5FeTi_3O_{15}$ (BTF) film at room temperature (RT),[14] which gives us a new way of achieving magnetoelectric coupling in ferroelectric materials without long-range magnetic ordering. However, the mechanism of ME switching in BTF has not been clarified, or rather the switching path of the electric and magnetic orders is still unknown. BTF has a structure with four-layered pseudoperovskite units of $(Bi_3Ti_3FeO_{13})^{2-}$, sandwiched by two $(Bi_2O_2)^{2+}$ slabs along the $c$-axis[15] (shown in Supplementary Figure S1). Thanks to its unique crystal structure, BTF has a high ferroelectric transition Curie point ($T_C$) of 750 °C (with a structural transition from the $A2_1am$ structure to the $I4/mmm$ structure) and large spontaneous polarization.[16, 17] The strong ferroelectricity is attributed to the rotation and tilt of the Ti(Fe)-O octahedra in the perovskite layers, where lone pair electrons of $Bi^{3+}$ ions induce Ti(Fe) ions to deviate from the centre of the Ti(Fe)-O octahedra along the $a$-axis,[18] while the $(Bi_2O_2)^{2+}$ slabs play an important role in space-charge compensation and insulation.[19, 20] Although BTF is attracting more attention as a promising candidate for single-phase RT multiferroics and thus for magnetoelectric coupling,[14, 21, 22] it is reported to have an antiferromagnetic ordering with a Néel transition temperature ($T_N$) of 80 K.[23] Therefore, it is

Ferroelectric and magnetic domain switching in BTF

obvious from the above reports that the origin of the magnetoelectric coupling at RT in BTF has yet to reach an agreement, let alone manipulation of nanoscale domains using electric field, magnetic field and external mechanical force.

In this work, $Bi_5FeTi_3O_{15}$ thin films were deposited on $Pt/TiO_2/SiO_2/Si$ substrates using pulsed laser deposition (PLD). An orthorhombic structure with space group *A2₁am* (36) was indexed by X-ray diffraction (XRD) as shown in Supporting Information Figure S1. The detailed microstructure and component analysis was carried out by transmission and scanning electron microscopy (TEM and SEM) (see Supporting Information Figure S2).We have visualized the switching of both ferroelectric and magnetic domains by electrical field and magnetic field using a SPM system. The time decay feature of magnetic domain switching accompanying ferroelectric domain switching is revealed after removing the electrical field, which is in good agreement with the nature of short-range magnetic ordering in BTF film. More importantly, we were successful in switching ferroelectric domains, as well as magnetic domains, by using a mechanical force applied on the probe tip of the SPM system. We believe that similar phenomena can be observed in other typical multiferroic materials. Therefore, our findings open up a new route to manipulate the magnetism in multiferroics by mechanical force and broaden the range of applications for future design of multifunctional devices made from multiferroic materials.

## EXPERIMENTAL PROCEDURES

### Film deposition

$Bi_5Ti_3FeO_{15}$ (BTF) thin film was deposited on $Pt/TiO_2/SiO_2/Si$ substrates by a pulsed laser deposition (PLD) system with the laser source at 355 nm and a repetition rate of 10 Hz. The ceramic target for BTF deposition was prepared using a conventional solid-state reaction process.

BTF thin films were deposited at 470~500 ºC for 1 hour, followed by an in-situ thermal annealing process for 30 min, and then cooled down to RT. The film thickness is ~269 nm (Figure S2).

**Macroscale electrical and magnetic measurements**

Pt top electrodes were coated on the surface of the BTF thin film through a shadow mask with a diameter of 100 μm to form a capacitor. Dynamic hysteresis measurements were carried out to verify the ferroelectricity using an aixACCT TF-1000 ferroelectric tester. (Figure 1 a,b,c). Magnetic properties of the as-deposited BTF film were measured using a superconducting quantum interference device (SQUID) (Quantum Design MPMS magnetometer). Here, the diamagnetic moment of the substrate was calculated from measurements on the bare substrate and subtracted from the raw data (Figure 1d).

**SPM measurement**

A Nanocute SPI 3800 SPM system and an E-sweep SPM system (Hitachi HiTech Science), which enable piezoresponse force microscopy (PFM), electrostatic force microscope (EFM) and magnetic force microscopy (MFM) measurements, were used to investigate local piezoelectric/ferroelectric and magnetic properties at RT. Figure S3 shows a schematic diagram of PFM and MFM operation. A conduction tip (Si cantilever coated with Rh) was used to apply electric field ($E$) and mechanical force ($F$) in contact mode PFM measurements of which the spring constant was 15 N/m, and the resonance frequency was 139 kHz. In PFM measurements, $A$cos images were recorded together with topographic images, applying 5 kHz and 2 V oscillation under a ±20 V dc bias. The $A$cos image correspondes to $A\cos(\theta + \omega t)$, where $A$ is the amplitude, $\theta$ is the phase shift between the driving voltage $V_{ac}$ and the voltage induced deformation, $\omega$ represents the frequency and $t$ is the sample thickness. After the electrical and mechanical poling, the VPFM and LPFM images were read with an ac voltage of 4 V and 5 kHz without applying any dc bias.   In

Ferroelectric and magnetic domain switching in BTF

order to reduce the effects of injected charges as much as possible during probing, we blew on the sample for 2 min using a static eliminator fan (HAKKO FE-510) between each writing process. The displacement-voltage curve was obtained directly by applying dc voltage on the film surface in PFM mode (Figure S4). Mechanical force was applied on the film surface by loading mechanical force on the tip during scanning. A magnetic tip coated with CoPtCr film was used in the MFM measurements. The reason for not using the MFM tip to do force scanning was to avoid the magnetization effect from the very weak magnetic field of the MFM tip. In MFM measurements, the trace is first recorded in tapping mode to image the topography of the surface; and the second step is in a lift mode in which the tip does not touch the film surface to assess the stray magnetic field perpendicular to the surface. To avoid the contribution of the interaction to the tip-sample force in MFM, [24] the tip-sample distance was kept at 10 nm to achieve a balance between avoiding the van der Waals contribution and improving the magnetic response. At this distance, the sample generates stronger stray fields, and the magnetic interaction force dominates, so the electrostatic force can be neglected. [25] Polarization reversal was also confirmed by scanning nonlinear dielectric microscopy (**SNDM**), as shown in supplementary Figure S6.

**RESULTS AND DISCUSSION**

**Domain switching by electric field**

Figure 2a-h are topographic images and corresponding vertical piezoresponse force microscope (VPFM), lateral piezoresponse force microscope (LPFM), MFM and EFM images before electrical switching. Figure 2i shows a box-in-box VPFM pattern of BTF thin film after electrically poling. The dark and   bright regions correspond to ferroelectric domain state with +P and –P, respectively, indicating that the as-deposited thin film can be reversibly switched between two ferroelectric state.

Ferroelectric and magnetic domain switching in BTF

The VPFM shows the information from out-of-plane (OP) polarization component, while the lateral piezoresponse force microscope (LPFM) provides information on the in-plane (IP) polarization component. Figure 2j shows a similar lithographic pattern, indicating that the perpendicular electric field could also partially switch the in-plane polarization. The VPFM and LPFM images present two components of the polarization vector,[26] corresponding to the orthogonal components to the *a*-axis of the as-deposited BTF thin film along the OP and IP directions. Since the ME coefficient in the BTF thin film is 400 mV/(Oe·cm) at RT,[14] it is possible to switch magnetic domains in this material by applying electric field (***E***). Figure 2k shows the corresponding MFM phase image in the same area after electric poling. A sharp contrast in the magnetic domains corresponding to the ferroelectric domains was observed, indicating that magnetic domains in BTF thin film were switched by ***E***. From the above measurements, we directly visualized the local magnetoelectric (ME) coupling in BTF thin film at RT. There are two mechanisms for ME coupling that could be considered: an ***E*** shifts the positions of the Fe-ions relative to the O-ions, changing the dipolar and exchange contributions to the magnetic interactions and modifying the electronic wave functions at the same time, which thus changes the magnetic coupling.[27] Another mechanism is that the ***E*** causes ferroelastic strain in the film, which could be mechanically transferred to the magnetic component, where it modifies the local magnetization of the domain structures.[28] In the pseudoperovskite layer of BTF thin film, there are two types of atomic position for Ti and/or Fe, one is a tetrahedral site and the other is an octahedral site. The occupancy of both sites by Fe or Ti atoms in the compounds is random and almost equal in general.[29] According to the Mössbauer spectra, there is no long range magnetic ordering in BTF.[14] However, weak magnetic hysteresis is observed at RT. For our sample, the shape of magnetic-temperature (MT) curve indicated a generic feature of magnetic disordered system due to random distribution of the Fe spins on available sites.[14] It is possibly due to Fe atoms randomly substitute Ti atoms in the pseudoperovskite layers of BTF. Generally, there is a charge disproportionation of Fe at octahedral coordinated site and the tetrahedral site.[29] Although the macroscopy analysis indicated the distribution of Fe and Ti atoms are random and almost equal in BTF,[30] a small inhomogeneous Fe distribution in a nano-region still persist, which will lead to a short range magnetic ordering in BTF thin film. An $E$ causes ferroelastic strain in the film and shifts the positions of the Fe-ions relative to the Ti and/or O-ions, changing the dipolar and exchange contributions to the magnetic interactions and modifying the electronic wave functions at the same time, which thus changes the magnetic coupling.[27] For multiaxial ferroelectrics as BTF materials, the ferroelectric domains are always coupled to ferroelastic domains in particular orientation. Once the ferroelectric domain is switched by $E$, the local ferroelastic strain state in the film is simultaneously rearranged. The ferroelastic domain is switched to a high-energy state accompanied with the polarization switching. Once $E$ was removed, a relaxation of the switched ferroelastic domain back to the original ferroelastic state will happen. The ferroelastic domains could either switch to the state antiparallizing to the ferroelectric polarization, or transited along a different distortion axis. Further investigation of the evolution of ferroelectric and magnetic domains after electric field switching was carried out. After removing the $E$ for 1 hour, no obvious change was observed in either of the VPFM and MFM images, (Figure 2m,o), but the switched pattern in the LPFM image became very weak (Figure 2n).  It is obvious that the relaxation of ferroelastic domain is anistropic, along the out-of-plane and in-plane directions, due to the inhomogeneous feroelastic strain distribution   in the BTF thin film. The IP ferroelastic relaxation is more effective and faster than OP. In an aurivillius layered structure such as BTF, there are four spontaneous polarizations along the <100> directions in the pseudoproveskite layer. Only the 180°

Ferroelectric and magnetic domain switching in BTF

ferroelectric domains are energetically stable while the 90° ferroelectric domains are eventually stablized by 90° ferroelastic domains.[31]   A clear switching pattern can still be observed in the VPFM image after removed the electric field for 22 hours (Figure 2q), although it was weakened somewhat due to the out-of- plane ferroelastic relaxation. However, almost no contrast was observed in the LPFM image (Figure 2r) after 22 hours, the ferroelastic domain relaxation alone IP direction was completed after 22 hours. Interestingly, a relaxation of magnetic domains also takes place as shown in Figure 2s, where the switched magnetic pattern nearly fades away after 22 hours. It is suggests that the *E*-switched magnetic domains are coupled to ferroelastic domains, which might be caused by the unique magnetic ordering status in the BTF thin film. It also reveals that the magnetic ordering is not coupling directly with ferroelectric ordering in BTF, but coupled though ferroelastic strain. The ferroelastic strain relaxation will destabilize the magnetoelastic coupling in the BTF thin film, thus, a time decay behavior of magnetic domain switching will be observed as shown in Figure 2.

During SPM imaging and quantification of local electrical properties of ferroelectric materials, there is an unavoidable issue that should be considered, that is, the effect of space charges introduced by the cantilever/tip. The resultant surface charge screening is a general feature which affects the electrostatic and transport properties of surfaces, and thus, could probably lead to an incorrect interpretation of the data during the SPM measurements.[32, 33] In general, electrostatic force microscope (EFM) is used to investigate the surface charge density ($\sigma_s$) of materials. The EFM image shown in Figure 2l was taken immediately after the electric poling, the bright contrast in the EFM pattern indicates the sign of the measuring image is not determined only by the polarization charges, but also uncompensated charges due to charge injection by electric poling. In this work, we don't apply dc voltage during EFM scanning, so that the EFM image arises only from variation in the absolute value of electric field above the sample surface.[39] The surface charge distribution is also recorded after remove the $E$ for 1hour (Figure 2p). The screening effect which due to the adsorption of charge species from the ambient significantly reduce the surface mobile charges once remove the $E$, and the charges are compensated completely after removed the $E$ for 2 hours (Figure 2t).

**Domain switching by mechanical force**

To investigate the influence of mechanical force on both the ferroelectric and magnetic domains of our BTF film, we have applied a set of vertical loading forces on the film. As shown in Figure 3a, the BTF thin film was firstly poled by +20 V in an area of $8 \times 8$ µm$^2$ to obtain an uniformly arranged "mono-domain"; then the sample was scanned under a series of loading forces of 14 nN, 50 nN, 100 nN, 500 nN, 1000 nN, and 1300 nN, respectively, in an area of $4 \times 4$ µm$^2$ in the centre; finally, the switched area was read by scanning a larger area of $12 \times 12$ µm$^2$ at 5 kHz in VPFM, LPFM and MFM mode. As shown in Figure 3a, for the VPFM images, when the tip force ($F$) is lower than 100 nN, no obvious switching is observed in the "force switching" area in the centre. Further increasing $F$, the contrast between the area that is switched by $E$ and $F$ becomes more obvious. As the tip force is increased to 1300 nN, although the brightest contrast is obtained, some nonswitched domains are still exit in the centre of VPFM indicating an incomplete polarization reversal in the mechanical force switched region. The LPFM images show different switching phenomenon compared with VPFM images: a bright pattern is obtained in the $E$-poled region which is in accordance with Figure 2, however, the loading force seems have little effect on the ferroelectric domains along IP direction because no obvious domain switching observed in the $F$-poled region in the centre; we consider that the sign of IP projection of the ferroelastic strain is opposite with the IP projection of piezostrain generated by external mechanical force and the magnitude is bigger than the latter, so that the back switching of ferroelastic domain becomes more obvious in the $E$-switched area. For the MFM images, the magnetic domain switching become more obvious with increasing the $F$. We applied the $E$ and $F$ repeatedly in the same area, the $E$-switched and $F$-switched pattern in MFM become more homogeneous when applied force is up to 1300nN. When the $F$ increased to 1300nN, the magnetic domains in the $F$-poled region were mainly switched, indicated that the magnetization could be controlled by external mechanical force. The increasing trend towards switching of magnetic domains with increasing the load infers an enhancement of magneto-elastic coupling with increasing tip stress in BTF thin film. We did not apply higher forces because a higher force would cause wear on the tip, and the image would become blurred. The surface potential ($\Delta V$) in PFM and magnetic response in MFM as a function of applying $F$ was recorded in Figure 3b, c, the results were acquired based on a mean square error for five times. A transition is observed in both PFM and MFM response when applying mechanical force of 100nN, infers that the domain switching energy barrier is around 100nN. According to the above-mentioned experimental results, both ferroelectric and magnetic domains are switched by $F$, indicating that the tip stress can induce not only local ferro-electro-elastic polarization switching but also ferroelastic-magnetic domain switching. We have to note that the polarization reversal could not be observed when applying mechanical force without an initial positive dc voltage poling. In addition, no domain switching were observed if the thin film is initially poled by negative voltage. The tip stress induced switching pattern as shown in Figure 3a is not as bright as the $E$-induced switching pattern in Figure 2, due to the ferroelastic domain generation during the strain relaxation when electric field and mechanical force in the film is removed.[34]

Ferroelectric and magnetic domain switching in BTF

A mechanical force applied on a ferroelectric material such as BTF will lead to several effects: (1) piezoelectricity is a linear electromechanical interaction describes the electricity related to the electric charge separation by stress; (2) ferroelasticity which is the phase change results in a spontaneous strain generated by the stress; and (3) flexoelectricity which describes a spontaneous polarization result from the strain gradient. In this work, the tip stress generated spontaneous strain from ferroelasticity and led to a charge separation and then generated an electric field ($E_f$), which was comparable with the coercive field, finally switched ferroelectric domains in the sample. For a thin film in nanoscale, the flexoelectricity could not be ignored, thus $E_f$ attributed from both piezoelectricity and flexoelectricity. The $E_f$ gives rise to a large field gradient divergence that persists in BTF film, eventually resulting in a polarization reversal. The preferred direction of the as-deposited BTF thin film [119] (supplementary Figure 1) is at an angle to the spontaneous polarization direction [100]. Since the mechnical force was applied on the tip along out-of-plane direction which was perpendicular to the film surface, the pizeoresponse along the out-of-plane direction in the thin film is stronger than the in-plane. Besides the piezoelectricity, tip stress modified elastic strain could also account for the switching of polarization from up to down. Due to the anisotropic elastic strain in BTF thin film, the strain relaxation along in-plane (IP) direction is faster than the out-of-plane (OP).  As shown in supplementary Figure S5, the ferroelectric domain and magnetic domain were subsequently switched by applying +20 V after the mechanical switching, indicating the mechanical force could act similarly as an electric field on domain switched ferroelectric domain and magnetic domain.

The mechanical force introduces a piezostrain in nano-region, which modifies the original elastic strain state generated from the thermal process during film deposition,[35] and results in a consequent reorientation of the local magnetism driven by the strain coupling between small magnetic volume

Ferroelectric and magnetic domain switching in BTF

and ferroelastic domians. [36] The switching process is illustrated in Figure 3d. We suppose that the rotational moment and polar distortions in BTF would be coupled under certain conditions involved in strain state, as in BiFeO$_3$.[25] The tip stress induced the $E_f$ changes the free energy profile asymmetrically, which would also take account of the factors for switching the domains. Further investigation on the coupling configuration between ferroelectric domains and magnetic domains would be complicated, it is necessary to discover the ME coupling mechanism in high quality BTF single crystals.

**Domain switching by magnetic field**

In order to investigate the effects of magnetic field on the domain switching in BTF thin film, an alternating magnetic field ($H$) was applied on the BTF thin film. Figure 4b-c shows MFM images of an as-deposited BTF thin film under the magnetic field of 0 mT, 50 mT, and 100 mT, respectively. Due to the weak magnetization of the BTF thin film, no magnetic domains could be observed without applying $H$. The magnetic domains became more obvious with increasing $H$, indicating the magnetic domains could be manipulated by $H$. Due to the limitations of dc bias in this system, the sample was poled by dc bias of ±10 V in VPFM mode, as shown in Figure 4f. Although the ferroelectric domains were not fully switched, the box-in-box PFM pattern was observed. The corresponding MFM image was scanned in the same area in vacuum ($7.1 \times 10^{-4}$ Pa) (Figure 4j). A clear switching pattern was observed, even the magnetic domains were not completely switched, and indicating the ME coupling in BTF is stable in vacuum at RT. An in-plane $H$ of 100 mT was applied on the sample (schematic diagram shown in Figure 4m) to investigate the effect of $H$ on the $E$-switched domains. PFM and MFM images were collected with shift the direction of $H$ as shown in Figure 4g,h, respectively. The ferroelectric domains were rarely affected by +100 mT, while the magnetic domains were enhanced by an $H$ of +100 mT,

Ferroelectric and magnetic domain switching in BTF

which indicated that the +100 mT field lied in accordance with the easy axis of BTF helps complete the switching of magnetic domains which were incompletely switched by **E**. When the in-plane **H** was applied in the reverse direction (marked as -100 mT), the magnetic domains were switched off (Figure 4l). However, some of the switched ferroelectric domains were switched back when the direction of the direction of **H** was changed to -100 mT (Figure 4h), and this result indicates that the **H** in this direction helps reducing the back switching energy of ferroelectric domains. Since we has discussed above that the back switching of ferroelectric domains is mainly due to the strain relaxation of ferroelastic domains, this phenomenon infers a unique magneto-elasto coupling behavior in BTF thin film, indicating that the coupling is highly depend on the symmetric orientation. The magnetization switching can also switch the ferroelectric polarization through ferroelastic strain. Reverse process could be also expected as follows: The magnetic $Fe^{3+}$ ordering is relative to the polarization in the BTF film. The **H** forces a rotation of $Fe^{3+}$ spins that leads to a symmetry distortion of the crystal structure, and changes the strain state in the film, thereby controls the polarization of the BTF film.

**Exclusion of screening charge effect**

Figure 5a shows a topographic image of the as-deposited BTF thin film. In order to exclude the effect of mobile charges on the PFM image, we carried out a series of experiments: at first, we switched the domains by the electric field and tip stress. Apparent switching is due to a combined effect of externally introduced mobile charges and polarization-generated screening charge, as shown in Figure 5b. Secondly, we scanned the switching area again without applying dc voltage. During the probing process, the top and bottom electrodes were short-circuited for a short time, as shown in Figure 5c. There was no contrast, as expected by the screening of the film surface charges, and thus, the electrostatic force on the cantilever does not show contrast on top of the up/down poled regions. Figure 5d presents the PFM image after removal of the short-circuit. The same domain structure as initially written is observed, indicating that the written domains are not affected by short-circuiting the sample surface and the bottom electrode through the cantilever and persist for a long time, which confirms the signal convolution of the up and down domains.

**CONCLUSION**

In summary, we have demonstrated the manipulation of both ferroelectric and magnetic domains by electric field, magnetic field and mechanical force in a BTF thin film at room temperature. However, due to the relaxation of ferroelastic domains in the BTF film and possible short-range magnetic ordering in Fe rich nano regions, the magneto-elastic coupling is not stable, *E*-induced magnetic domain switching is fade away after the electric field is removed. While the successfully manipulation of both ferroelectric domain and magnetic domain by mechanical force, proved the switching path role of ferroelastic strain in the magnetoelectric coupling in BTF thin film, which would be important for further understanding the magneto-elastic-electric coupling in multiferroics. The mechanical switching of both ferroelectric and magnetic domain behavior in multiferroics shows a complex intercoupling among polarization, magnetization, and strain, provides the possibility of manipulating one physical parameter by using one of the three external stimuli: magnetic field, electrical field, and mechanical force, which could yield an additional degree of freedom in novel multifunctional device design. The mechanical switching of both ferroelectric domains and magnetic domains gives rise to an enormous potential application in high-density data storage via mechanical means, might help to overcome the leakage or dielectric breakdown as in direct electric switching.

**References**

Ferroelectric and magnetic domain switching in BTF

**Figures**

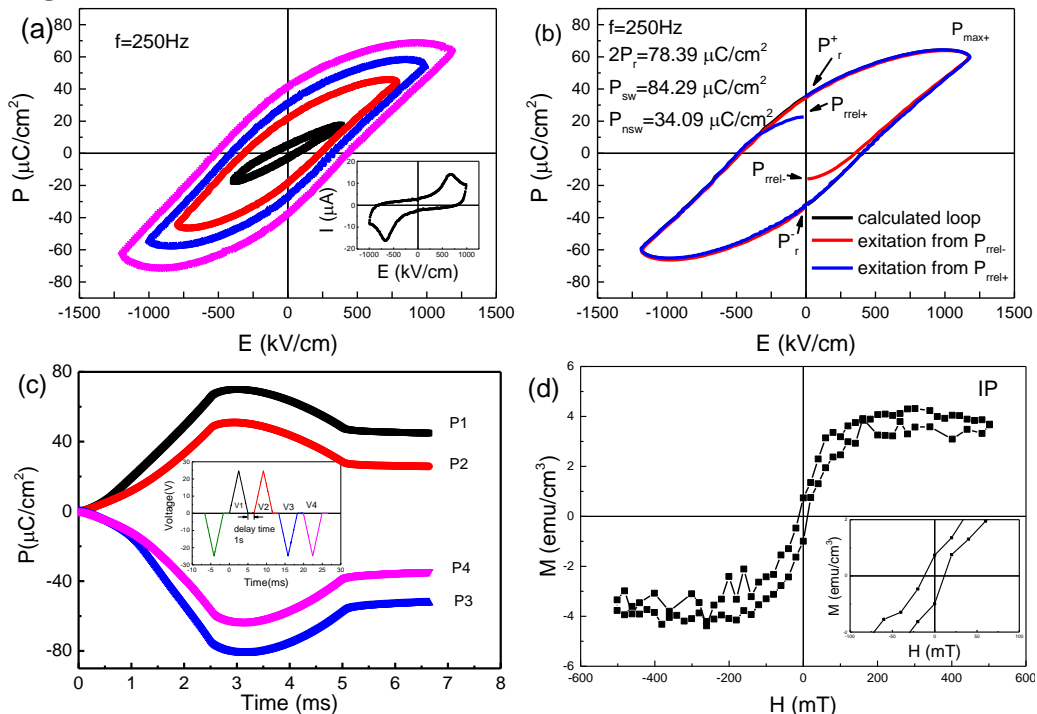

**Figure 1. Ferroelectric dynamic hysteresis measurement and magnetic hysteresis loop.** (a) Electric field dependence of ferroelectric polarization (*P-E* loops) of BTF thin film measured at 250 Hz, inset shows the correspond I-V curve; (b) a giant remanent polarization of

$$2P_r = \left| P_r^+ \right| + \left| P_r^- \right| = 78.39 \ \mu C/cm^2$$ is obtained by bipolar excitation hysteresis measurement at 250

Ferroelectric and magnetic domain switching in BTF

Hz. Triangular voltage excitation was applied on the sample. The red loop start from negative relaxed remanent polarization state ($P_{rrel-}$) then goes to the positive saturation ($P_{max+}$) and end at the positive remanent polarization state ($P_r^+$). The blue loop start from the positive relaxed remanent polarization state ($P_{rrel+}$) and turns into the negative remanent polarization state ($P_r^-$). The switching polarization $P_{sw} = P_{max+} - P_{rrel-}$, and the non-switching polarization $P_{nsw} = P_{max+} - P_{rrel+}$. (c) PUND switching polarization as a function of voltage, inset is the waveform of the applied triangle pulse: the write pulse rise-time is 2.5 ms, and the read pulse delay is 1 s; (d) in-plane (IP) magnetic hysteresis (*M-H*) loop measured at room temperature (RT), with the inset showing the enlarged central part of the *M-H* loop of the BTF thin film.

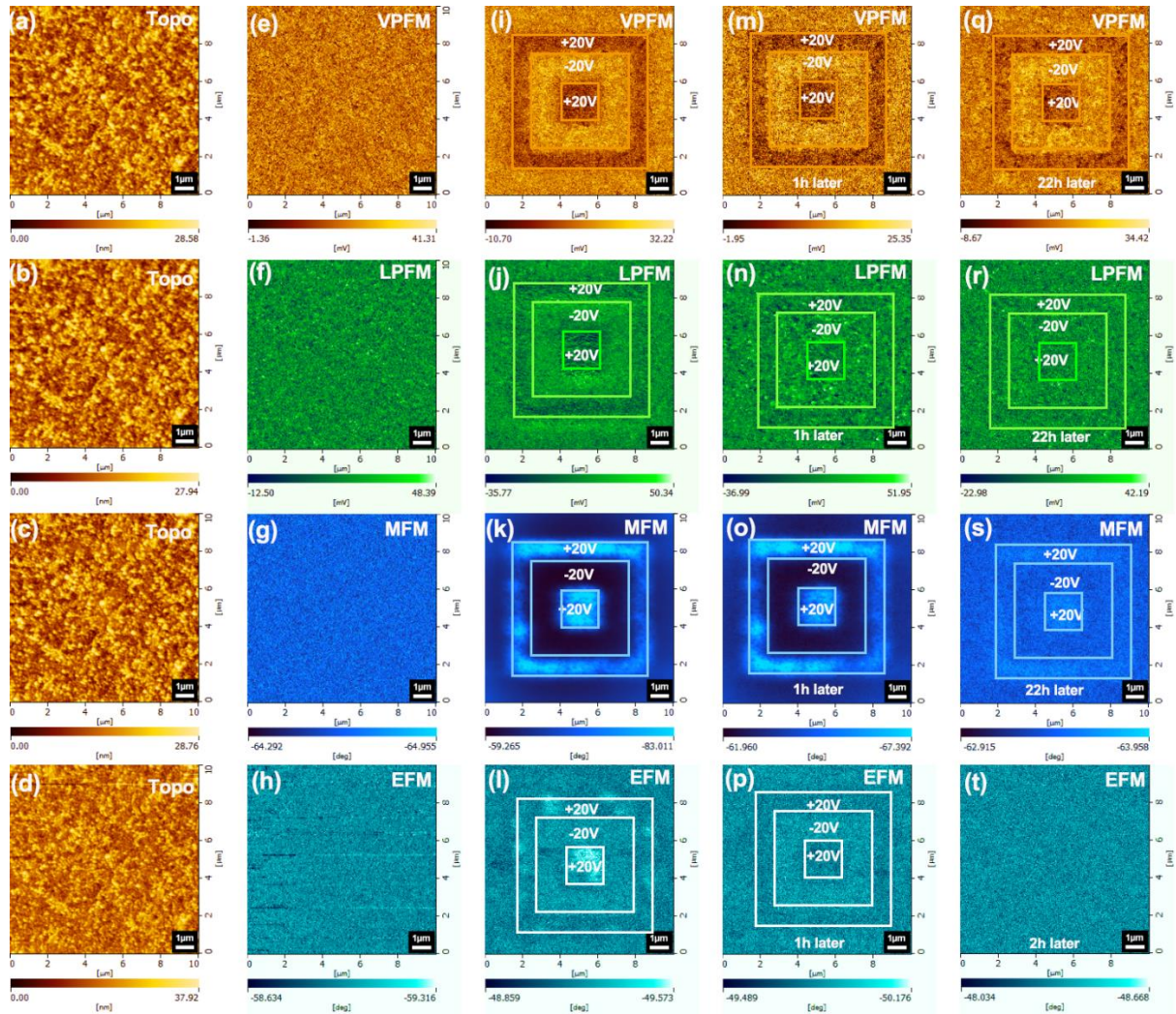

**Figure 2.** Reversible control of both ferroelectric and magnetic domains by electric field (E). (a-d) Topographic images, and (e-h) vertical piezoresponse force microscope (VPFM) image, lateral piezoresponse force microscope (LPFM) image, magnetic force microscope (MFM) phase image and corresponding electrostatic force microscope (EFM) image of the BTF thin film before electric poling, (i) VPFM image poled by +20 V in the area of 7×7 µm², -20 V in 5×5 µm², and +20 V in 2×2 µm², with reading over 10×10 µm², (j) corresponding lateral piezoresponse force microscope (LPFM) image, and (k) magnetic force microscope (MFM)

phase image, and (l) electrostatic force microscope (EFM). (m-t) Evolution of VPFM, LPFM, MFM, and EFM images of the poled BTF thin film collected after removing the electric field.

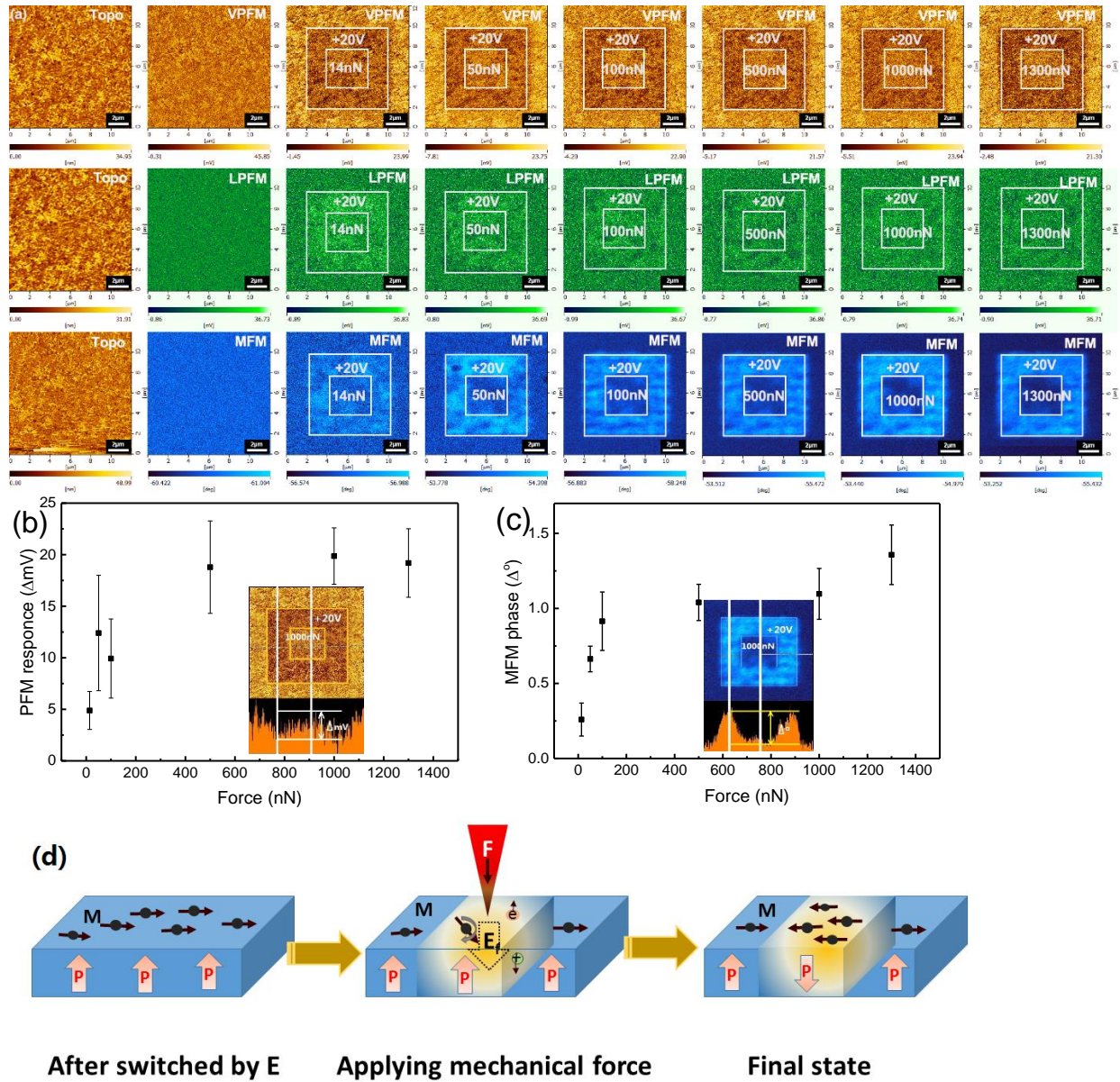

**Figure 3.** Mechanical force induced ferroelectric and magnetic domains switching. (a)topographic images and corresponding VPFM, LPFM and MFM images before and after electric and mechanical switching. The BTF thin film which was firstly poled by positive electric bias of +20 V over 8 × 8 μm², and then switched by mechanical forces of 14 nN, 50 nN, 100 nN, 500 nN, 1000 nN, and 1300 nN in the center over 4 × 4 μm², respectively; (b) the surface potential (ΔV) as a function of F; (c) magnetic response (Δº) as a function of F; (d) schematic illustration of ferroelectric domain switching and magnetic domain switching when applying mechanical force.

Ferroelectric and magnetic domain switching in BTF

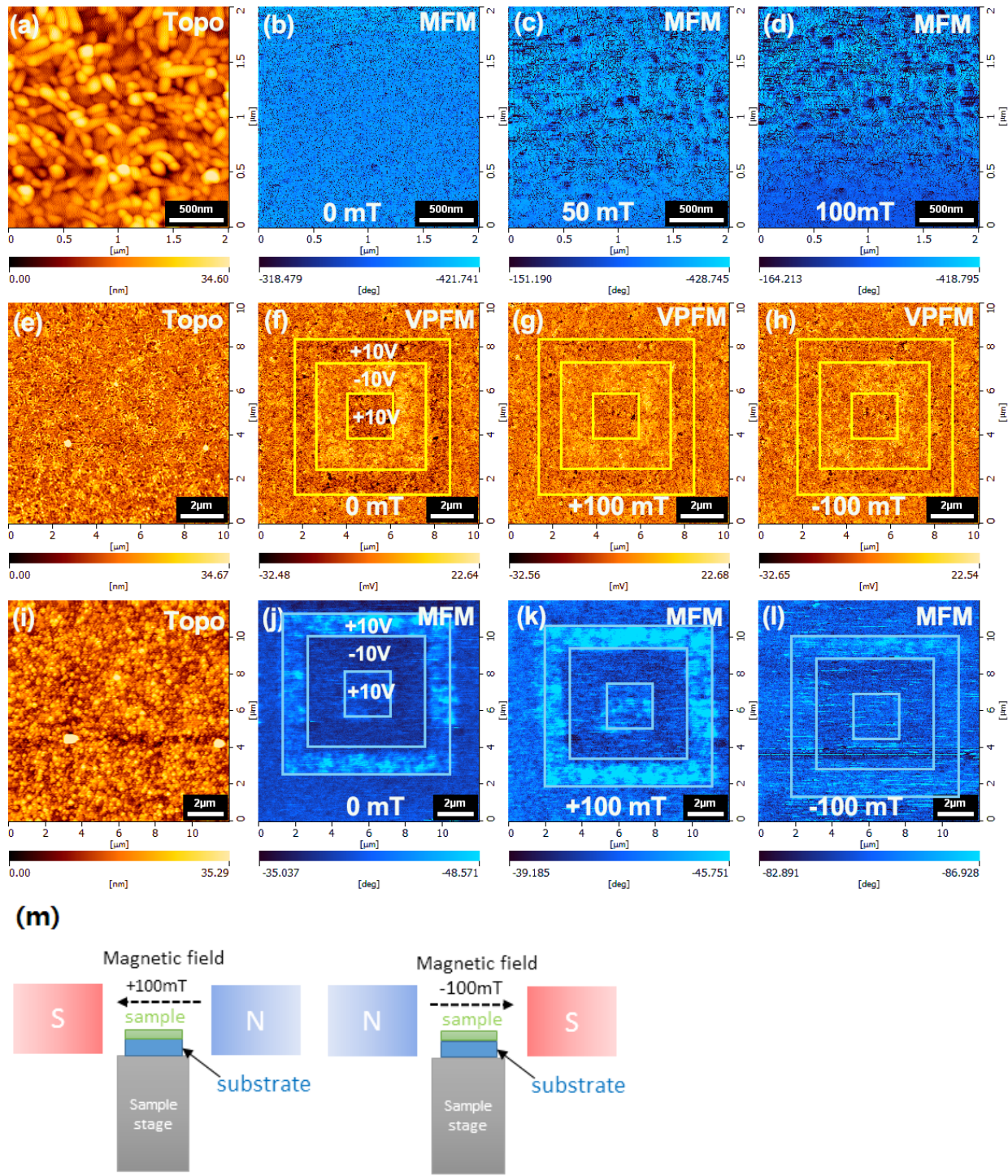

**Figure 4.** Control of both ferroelectric and magnetic domains by magnetic field (**H**). topography (a) and MFM images measured under magnetic field of (b) 0 mT, (c) 50 mT, and (d) 100 mT, with the part in the white ellipse showing a typical phenomenon: the magnetic domains become more apparent with increasing magnetic field; (e) topography and (f) VPFM image of BTF thin film which was firstly poled by electric field; corresponding (i) topography and (j) MFM image

Ferroelectric and magnetic domain switching in BTF

measured in vacuum. (g) VPFM image and (k) MFM image measured under magnetic field of +100 mT. (h) VPFM image and (l) MFM image under -100 mT. (m) Schematic diagram of in-plane magnetic field applied on the BTF thin film. The signs "+" and "-" for the magnetic field represent different magnetic field direction.

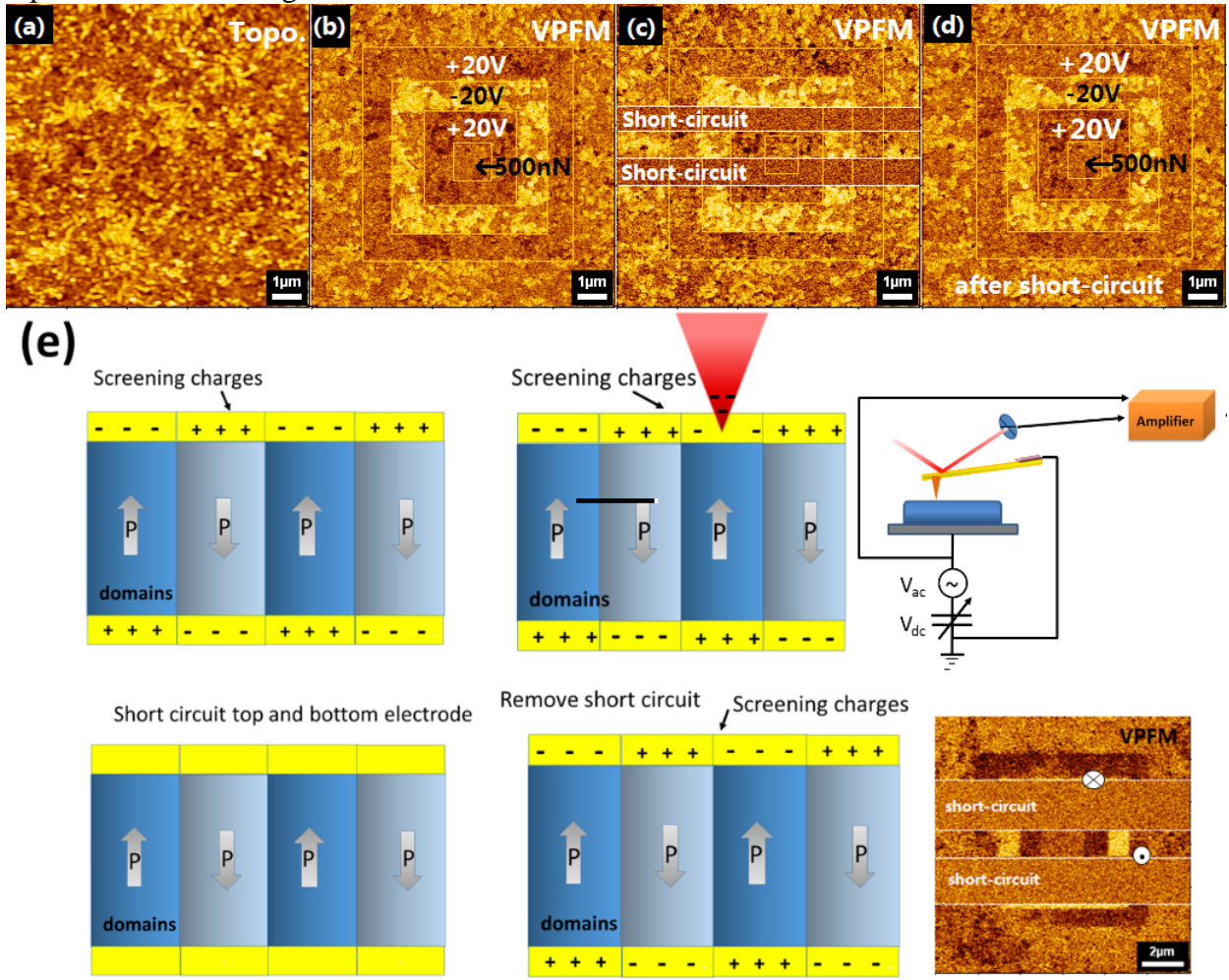

**Figure 5.** Exclusion of screening charge effect. (a) Topographic image of the BTF film; (b) corresponding VPFM box-in-box pattern written by ±20 V and 500 nN, (c) short-circuited top and bottom electrodes during scan, (d) VPFM image scanned after removing the short circuit, (e) schematic illustrations of charge screening behavior before and after short-circuiting the top and bottom electrodes.